\documentclass[]{mn2e}
\usepackage{graphicx}
\usepackage{epsfig}

\newcommand \cm           {\,{\rm cm}}

\newcommand \g            {\,{\rm g}}
\newcommand \K            {\,{\rm K}}

\newcommand \Qabs         {Q_{\rm abs}}

\newcommand \mum          {\,{\rm \mu m}}

\newcommand \simali       {\sim\,}

\newcommand \rc          {r_{\rm compact}}
\newcommand \rp          {r_{\rm porous}}
\newcommand \olv          {\rm MgFeSiO_4}
\newcommand       \simlt        {\leq}
\newcommand       \simgt        {\geq}

\title[On the Anomalous Silicate Emission Features of AGNs]
      {On the Anomalous Silicate Emission Features of AGNs:\\
       A Possible Interpretation Based on Porous Dust}

\author[M.P.~Li, Q.J.~Shi \& A.~Li]
       {M.P.~Li$^{1}$, Q.J.~Shi$^{2}$ and Aigen Li$^{1}$
       \thanks{%
                E-mail: limo@missouri.edu,
                        qingjiongshi@gmail.com,
                        lia@missouri.edu
                }\\
  $^1$Department of Physics \& Astronomy, 
      University of Missouri, Columbia, MO 65211, USA\\
  $^2$Department of Science and Engineering
      of Shuda College, Hunan Normal University,
      Changsha, Hunan 410081, China
      }
\begin{document}
\date{Received date  / Accepted date }
\pagerange{\pageref{firstpage}--\pageref{lastpage}} \pubyear{2008}

\maketitle

\label{firstpage}
\begin{abstract}
The recent {\it Spitzer} detections of the 9.7$\mum$ Si--O
silicate emission in type 1 AGNs provide support for 
the AGN unification scheme. 
The properties of the silicate dust are of key importance to 
understanding the physical, chemical and evolutionary 
properties of the obscuring dusty torus around AGNs. 
Compared to that of the Galactic interstellar medium (ISM),
the 10$\mum$ silicate emission profile of type 1 AGNs 
is broadened and has a clear shift of peak position to 
longer wavelengths.
In literature this is generally interpreted as
an indication of the deviations of the silicate 
composition, size, and degree of crystallization
of AGNs from that of the Galactic ISM.
In this {\it Letter} we show that the observed peak shift and
profile broadening of the 9.7$\mum$ silicate emission feature 
can be explained in terms of porous composite dust
consisting of ordinary interstellar amorphous silicate, 
amorphous carbon and vacuum.
Porous dust is naturally expected 
in the dense circumnuclear region around AGNs,
as a consequence of grain coagulation.
\end{abstract}


\section{Introduction}
Dust is the cornerstone of the unification theory
of active galactic nuclei (AGNs). This theory proposes
that all AGNs are essentially ``born equal'' --
all types of AGNs are surrounded by an optically thick
dust torus and are basically the same object
but viewed from different lines of sight
(see e.g. Antonucci 1993; Urry \& Padovani 1995):
type 1 AGNs, which always display broad hydrogen emission lines 
in the optical and have no obvious obscuring effect, 
are viewed face-on which allows a direct view 
of the central nuclei, while type 2 AGNs are viewed 
edge-on with most of the central engine and broad
line regions being hidden by the obscuring dust.

Silicate dust, a major solid species in the Galactic
interstellar medium (ISM) as revealed by the strong
9.7$\mum$ and 18$\mum$ bands 
(respectively ascribed to the Si--O stretching 
and O--Si--O bending modes in some form of silicate material,
e.g. olivine Mg$_{2x}$Fe$_{2-2x}$SiO$_4$),
has also been detected in AGNs 
both in {\it emission} and in {\it absorption}
(see Li 2007 for a review).

The first detection of the silicate {\it absorption} 
feature in AGNs was made at 9.7$\mum$ for 
the prototypical Seyfert 2 galaxy NGC\,1068
(Rieke \& Low 1975; Kleinmann et al.\ 1976),
indicating the presence of a large column of silicate dust 
in the line-of-sight to the nucleus.
It is known now that most of the type 2 AGNs display 
silicate {\it absorption} bands 
(e.g. see Roche et al.\ 1991,
Siebenmorgen et al.\ 2004, 
Hao et al.\ 2007, Spoon et al.\ 2007, Roche et al.\ 2007)
which is expected from the AGN unified theory
-- for a centrally heated optically thick torus viewed 
edge-on, the silicate features should be in absorption.

For type 1 AGNs viewed face-on, one would expect to see 
the silicate features in {\it emission} since the silicate dust 
in the surface of the inner torus wall will be heated to 
temperatures of several hundred kelvin to $\simali$1000$\K$
by the radiation from the central engine, 
allowing for a direct detection of the 9.7$\mum$ 
and 18$\mum$ silicate bands emitted from this hot dust. 
However, their detection has only very recently 
been reported in a number of type 1 AGNs
covering a broad luminosity range,
thanks to {\it Spitzer}
(Hao et al.\ 2005, Siebenmorgen et al.\ 2005, 
Sturm et al.\ 2005, Weedman et al.\ 2005, 
Shi et al.\ 2006, Schweitzer et al.\ 2008).
It is worth noting that the silicate emission 
features have recently also been detected in
type 2 QSOs (Sturm et al.\ 2006, Teplitz et al.\ 2006).
%

\begin{figure}
\begin{center}
\psfig{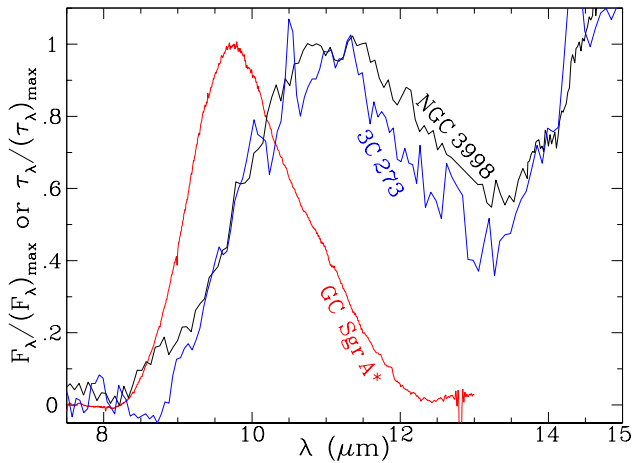}
\end{center}
\caption{\label{fig:sil_obs_spec}
         Comparison of the silicate emission features of
         the quasar 3C\,273 (Hao et al.\ 2005)
         and the low-luminosity AGN NGC\,3998 (Sturm et al.\ 2005)
         with the silicate absorption feature 
         of the ISM toward the Galactic 
         center object Sgr A$^{\ast}$ (Kemper et al.\ 2004).
         Most notably is the long wavelength shift of the
         peak positions of 3C\,273 and NGC\,3998 in comparison
         with the ISM profile.  
         All profiles are normalized to their peak values. 
         }     
\end{figure}

Compared to that of the Galactic ISM,
the 9.7$\mum$ silicate emission profiles of 
some AGNs appear ``anomalous''.
As illustrated in Figure \ref{fig:sil_obs_spec},
both quasars (high luminosity counterparts of 
Seyfert 1 galaxies; Hao et al.\ 2005, 
Siebenmorgen et al.\ 2005)
and the low-luminosity AGN NGC\,3998 
(Sturm et al.\ 2005)
exhibit silicate emission peaks
at a much longer wavelength ($\simali$10--11.5$\mum$),
inconsistent with the ``standard'' silicate ISM dust
(which peaks at $\simali$9.7$\mum$).
The 9.7$\mum$ feature of NGC\,3998 is also much broader
than that of the Galactic ISM (Sturm et al.\ 2005).\footnote{%
  We should note that the width of the interstellar 9.7$\mum$
  Si--O absorption feature is not universal, 
  but varies from one sightline to another (see Draine 2003).
  Generally speaking, it is relatively narrow in
  diffuse clouds and broad in molecular clouds
  (Bowey et al.\ 1998).
  In contrast, its peak wavelength is relatively stable.
  }

The deviations of the silicate emission profiles of
type 1 AGNs from that of the Galactic ISM dust 
are generally interpreted as an indication of 
the differences between AGNs and the Galactic ISM
in the composition, size distribution, and degree 
of crystallization of the silicate dust (Sturm et al.\ 2005).
However, we show in this {\it Letter} that the observed 
peak shift and profile broadening of the 9.7$\mum$ silicate 
emission features of AGNs can be explained in terms of 
porous composite dust consisting of ordinary interstellar
amorphous silicate, amorphous carbon\footnote{%
  There must exist a population of carbonaceous dust
  in the AGN torus, as revealed by the detection of
  the 3.4$\mum$ absorption feature, attributed to 
  the C--H stretching mode in saturated aliphatic 
  hydrocarbon dust (see Li 2007 for a review).
  Whether the bulk form of the carbonaceous component
  in AGNs is amorphous carbon or hydrogenated amorphous
  carbon is not clear. 
  }
and vacuum, without invoking an exotic dust composition 
or size distribution. 
In the dense circumnuclear region around AGNs,
a porous structure is naturally expected for the dust 
formed through the coagulation of small silicate
and carbonaceous grains.

\section{Model of porous composite dust
         \label{sec:model}}
To model the porous composite dust, we adopt the multi-layered sphere
model originally developed by Voshchinnikov \& Mathis (1999).
This model assumes that the dust consists of many concentric 
spherical layers of different types of materials.
Each of the material has a given volume fraction.
Voshchinnikov et al.\ (2005, 2006) demonstrated that 
the optical properties of porous dust calculated from
the multi-layered sphere model is in close agreement 
with that from the discrete dipole approximation
(DDA; Draine 1988).While the DDA method is computationally
very time-consuming, the multi-layered sphere model is
computationally much less demanding but still accurate
for computing the integral scattering characteristics 
(e.g. extinction, scattering, absorption cross sections,
albedo and asymmetry parameter) 
and becomes a robust tool to model the optical properties
of composite dust with a porous structure.

We consider three types of dust materials in the model:
amorphous silicate with olivine composition ($\olv$), 
amorphous carbon, and vacuum.
We take the optical constants of amorphous olivine $\olv$
from Dorschner et al.\ (1995); 
for amorphous carbon, we take those of Rouleau \& Martin (1991). 

In the multi-layered sphere model, we need to specify
(1) $n_{\rm layer}$ -- the number of layers;
(2) $P$ -- the dust porosity (i.e. the volume fraction of 
    vacuum in a fluffy porous grain);
(3) $m_{\rm carb}/m_{\rm sil}$ -- the mass ratio
    of amorphous carbon to amorphous silicate in a grain;
(4) $r_{\rm compact}$ -- the radius of the mass-equivalent 
    compact sphere.
Voshchinnikov et al.\ (2005) found that the optical 
properties of layered dust are independent of the number 
and position of layers when the number of layers 
$n_{\rm layer}$ exceeds 15. 
We therefore take $n_{\rm layer}=18$.\footnote{%
  This means that each layered dust has 6 shells,
  and every shell has 3 layers consisting
  of amorphous silicate, amorphous carbon and vacuum.
  }
We consider a range of porosities, with $P$
ranging from $P=0$ (compact dust) to $P=0.9$.
We take the mass ratio of amorphous carbon to 
amorphous silicate to be $m_{\rm carb}/m_{\rm sil} = 0.7$
which is estimated from the cosmic abundance constraints
(see Li \& Lunine 2003).\footnote{%
  With the mass density of amorphous silicate 
  $\rho_{\rm sil}=3.5\g\cm^{-3}$
  and the mass density of amorphous carbon 
  $\rho_{\rm carb}=1.8\g\cm^{-3}$, 
  this mass ratio corresponds to
  a volume ratio of 
  $V_{\rm carb}/V_{\rm sil}\approx 1.4$.
  }
We take the radius of the mass-equivalent 
compact spherical grain to be $\rc=0.1\mum$,
a typical size for interstellar dust.\footnote{%
  With a larger $\rc$, the peak wavelength $\lambda_{\rm peak}$
  and width of the 9.7$\mum$ Si--O band will be further
  red-shifted and broadened. 
  But as $\rc$ exceeds $\simali$0.5$\mum$,
  the 9.7$\mum$ Si--O feature fades away for 
  porous dust with $P>0.5$.
  }  
For a porous dust of porosity $P$ with the same mass
as that of the compact dust of radius $\rc$, 
its radius is $\rp = \rc/(1-P)^{1/3}$.
For a given $\rc$, $\rp$ moderately increases with $P$.

\begin{figure}
\begin{center}
\psfig{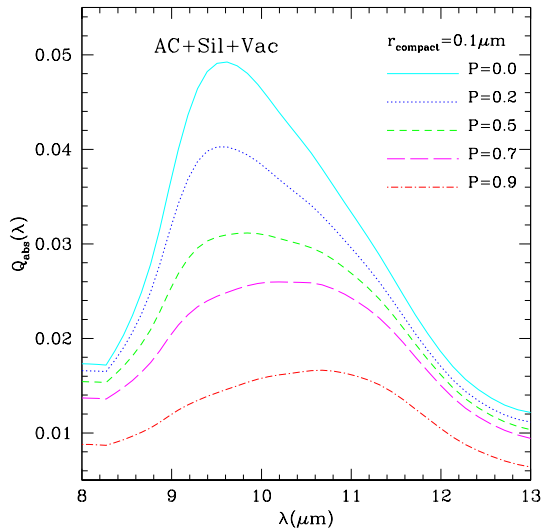}
\end{center}
\caption{
         \label{fig:Qabs_si_ac}
         8--13$\mum$ absorption efficiencies
         $\Qabs(\lambda)$ of porous composite grains
         consisting of amorphous silicate, 
         amorphous carbon and vacuum
         with $P =0, 0.2, 0.5, 0.7, 0.9$.
         Note the progressive broadening of 
         the 9.7$\mum$ Si--O feature and the progressive
         shift to longer wavelengths of its peak position 
         with the increase of the dust porosity $P$.
         }
\end{figure}

\section{Results\label{sec:results}}
Using the computational techniques of the multi-layered 
sphere model of Voshchinnikov \& Mathis (1999)
and assuming $n_{\rm layer}=18$,
$m_{\rm carb}/m_{\rm sil}=0.7$, and $\rc=0.1\mum$ (see \S2),
we calculate the absorption efficiency factors $\Qabs(\lambda)$ 
for porous composite dust of a range of porosities.
We show in Figure \ref{fig:Qabs_si_ac} the 8--13$\mum$
absorption efficiency $\Qabs(\lambda)$ calculated 
for porous composite dust consisting of amorphous
silicate, amorphous carbon and vacuum
with $P =0, 0.2, 0.5, 0.7, 0.9$.
With $\rc=0.1\mum$, the radius of the porous dust
corresponds to $\rp \approx 0.108, 0.126, 0.149, 0.215\mum$
for $P = 0.2, 0.5, 0.7, 0.9$, respectively
(for compact dust, the 9.7$\mum$ silicate absorption profiles 
are essentially identical for grains of radii
$r = 0.1$, 0.108, 0.126, 0.149, 0.215$\mum$, as expected 
since they are in the Rayleigh regime at $\lambda=10\mum$).
Most notably in Figure \ref{fig:Qabs_si_ac} are the progressive 
broadening of the 9.7$\mum$ Si--O feature and the progressive
shift to longer wavelengths of the peak position of this feature 
as the dust porosity $P$ increases:
while the silicate feature peaks at $\simali$9.6$\mum$ 
and has a FWHM (full width half maximum) $\simali$2.1$\mum$
for compact dust ($P=0$), its peak shifts to $\simali$10.6$\mum$ 
and its FWHM is broadened to $\simali$2.8$\mum$
for very fluffy dust ($P=0.9$); 
the 9.7$\mum$ Si--O feature is significantly flattened
as the porosity increases from $P=0$ to $P=0.9$.

\begin{figure}
\begin{center}
\psfig{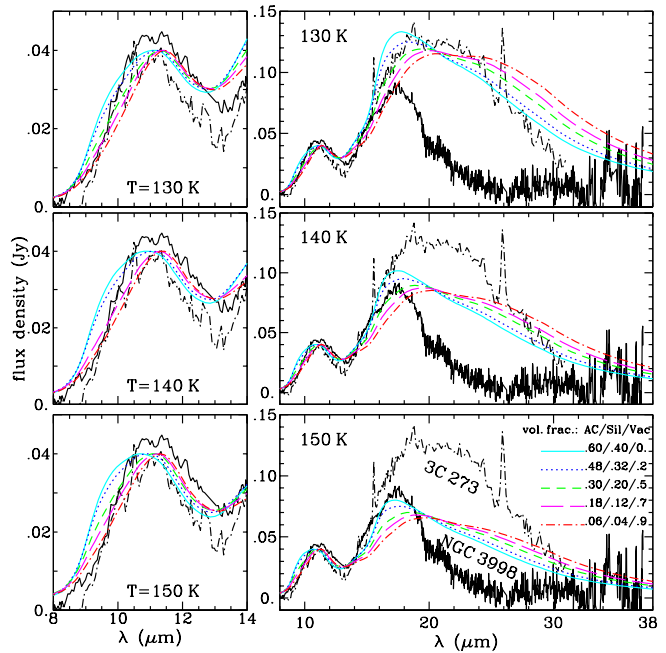}
\end{center}
\caption{
         \label{fig:agn_sil_irem_mod}
         Comparison of the silicate emission spectra 
         calculated from the porous composite dust model 
         with that observed in 3C\,273 (a bright quasar)
         and NGC\,3998 (a low luminosity galaxy).
         The model emission spectra are obtained by
         folding the absorption efficiencies $\Qabs(\lambda)$
         of porous composite dust 
         (see Fig.\,\ref{fig:Qabs_si_ac}) 
         with a Planck function at temperature $T$
         and then scaled to the flux densities
         of the 9.7$\mum$ features of 3C\,273 and NGC\,3998. 
         }
\end{figure}

\begin{figure}
\begin{center}
\psfig{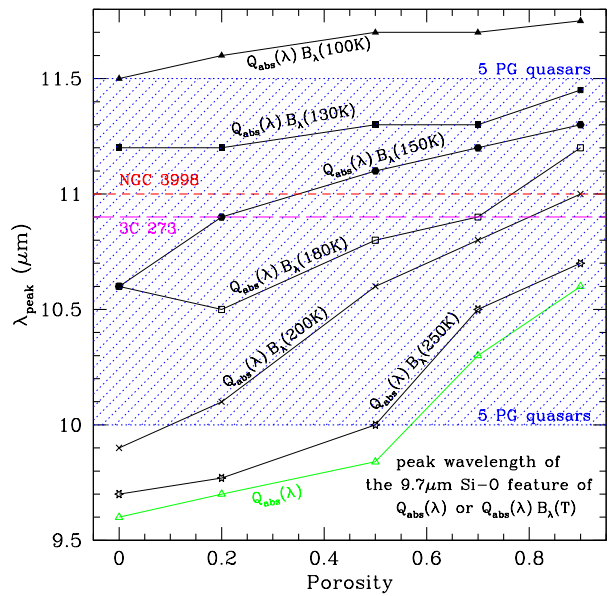}
\end{center}
\caption{
         \label{fig:agn_lambdap_porosity}
         Peak wavelengths of the ``9.7$\mum$'' Si--O
         silicate features of $\Qabs(\lambda)$ or 
         $\Qabs(\lambda)\,B_{\lambda}(T)$ 
         as a function of dust porosity for 
         a range of temperatures 
         $T$\,=\,100, 130, 150, 180, 200 and 250$\K$. 
         Also plotted are the peak wavelengths of 
         the 9.7$\mum$ feature of NGC\,3998 and 3C\,273.
         The shaded area plots the range of the peak wavelengths
         of the 9.7$\mum$ Si--O emission features
         of the five quasars observed by Hao et al.\ (2005):
         $10\mum \simlt \lambda_{\rm peak} \simlt 11.5\mum$.
         }
\end{figure}

In order to have a direct comparison between the porous dust
model with the ``anomalous'' silicate {\it emission} features 
observed in AGNs, we first multiply the calculated absorption efficiency
$\Qabs(\lambda)$ with a Planck function $B_\lambda(T)$ 
at temperature $T$. For a given $T$ we further
multiply the product of $\Qabs(\lambda)$ and $B_\lambda(T)$ 
with a constant to force the model to fit the observed
flux density of the 9.7$\mum$ feature.
This approach is valid if the silicate feature 
emitting regions are optically thin.\footnote{%
  The optical thin assumption is reasonable 
  if the silicate emission mainly comes from
  the unblocked surface layer of the inner torus wall.
  However, the inferred temperatures ($T<220\K$) for
  3C\,273 and NGC\,3998 
  (see Figs.\,\ref{fig:agn_sil_irem_mod},\ref{fig:agn_lambdap_porosity}) 
  are much lower than that expected for the dust
  near the surface of the inner torus wall
  where the temperature of silicate dust 
  should be close to its sublimation temperature
  ($\simali$800--1500$\K$; Kimura et al.\ 2002). 
  Sturm et al.\ (2005) suggested that this emission
  may actually originate from the extended narrow-line 
  regions (NLRs) which are beyond the torus.
  The NLR origin of the silicate emission 
  is consistent with the detection of silicate emission
  in type 2 QSOs (see Efstathiou 2006, Marshall et al.\ 2007,
  Schweitzer et al.\ 2008).
  Alternatively, the silicate emission could originate
  from the inner regions of the torus provided that the torus 
  is clumpy so that the light from the central engine
  is substantially attenuated
  (e.g. see Levenson et al.\ 2007,
   Nenkova et al.\ 2002, 2008).
  }

We take 3C\,273 (a bright quasar; Hao et al.\ 2005) 
and NGC\,3998 (a low luminosity AGN whose luminosity is 
$\simali$4--5 orders of magnitude below those of quasars;
Sturm et al.\ 2005) as two test cases.
While their 9.7$\mum$ Si--O features are virtually identical
(see Fig.\,\ref{fig:sil_obs_spec}),
at the 18$\mum$ O--Si--O feature region
they deviate significantly from each other 
(see Fig.\,\ref{fig:agn_sil_irem_mod}).

As shown in Figure \ref{fig:agn_sil_irem_mod},
with $T$\,$\simali$120--220$\K$, 
the porous composite dust model fits the broadened 
and long wavelength-shifted 9.7$\mum$ feature of 
3C\,273 and NGC\,3998 reasonably well. 
The required temperature ($T$\,$\simali$120--220$\K$)
is consistent with that constrained by Hao et al.\ (2005)
and Sturm et al.\ (2005). 
With $T=130\K$ the model (with $P \simlt 0.2$) is also 
in a reasonably good agreement with
the long-wavelength wing of the 18$\mum$ feature of 3C\,273. 
It is more challenging to fit that of NGC\,3998 
which is much weaker than that of 3C\,273
and peaks at $\simali$18.5$\mum$ 
(while the ``18$\mum$'' O--Si--O feature of 3C\,273
peaks at $\simali$20$\mum$).\footnote{%
  It is possible that the composition of 
  the silicate dust in NGC\,3998
  may differ from that in 3C\,273,
  suggesting that there might be
  significant environmental variations 
  (after all, NGC\,3998 is a LINER galaxy
  with an AGN luminosity $\simali$$1.7\times 10^4$
  times below that of the bright quasar 3C\,373;
  it is therefore possible that the silicate dust
  in NGC\,3998 subjects to different degrees
  of processing compared to that in 3C\,273.)
  As shown experimentally in Dorschner et al.\ (1995),
  the peak wavelength, width and strength 
  of the ``18$\mum$'' O--Si--O feature 
  (relative to the 9.7$\mum$ Si--O feature) vary 
  among silicate minerals of different composition.
  The optical constants of amorphous olivine $\olv$
  chosen here may not be the most suitable ones for NGC\,3998
  (e.g. the fast declining red wing of the 18$\mum$
   feature may suggest the presence of clino-pyroxenes
   [Wooden et al.\ 1999]).
  }
Qualitatively speaking, one requires 
a higher $T$ and a high porosity ($P\simgt 0.7$)
to fit the 18$\mum$ feature of NGC\,3998.\footnote{%
   One may ask why our models imply that the dust in 3C\,273 
   is cooler than the dust in NGC\,3998 while 3C\,273 is much 
   more luminous. A plausible answer is that the silicate emission
   actually originates from the NLRs far away from the central heating 
   regions or from a clumpy, attenuated torus.
   }
We should emphasize that the major purpose of 
this work is not to provide a detailed modeling 
of the silicate emission spectra of 3C\,273 and NGC\,3998,
but to put forward a hypothesis that the peak redshift
and profile broadening of the 9.7$\mum$ Si--O emission 
features observed in some AGNs could result from 
the porous structure of the dust.

In the Hao et al.\ (2005) sample, four quasars
(among the five PG quasars displaying the 9.7$\mum$ 
and 18$\mum$ silicate emission features, apart from 3C\,273) 
exhibit an 18$\mum$ O--Si--O emission feature 
intermediate between that of NGC\,3998 and that of 3C\,273
(while their 9.7$\mum$ features are all similar).
As can be seen in Figure \ref{fig:agn_sil_irem_mod}, 
the porous composite dust model 
with $T$\,$\simali$130--150$\K$ should be able 
to simultaneously fit both the 9.7$\mum$ features 
and the 18$\mum$ features of those quasars.

\section{Discussion\label{sec:discussion}}
It is well-known theoretically that as the size of
a silicate grain increases, the 9.7$\mum$ Si--O feature 
becomes wider with its peak shifted to longer wavelengths
(see Fig.\,9 of Dorschner et al.\ 1995, 
Fig.\,1 of Voshchinnikov \& Henning 2008).
But for compact silicate spheres to have its
9.7$\mum$ Si--O feature peaking at 
$\lambda_{\rm peak} >10.5\mum$, 
their size needs to exceed $\simali$2$\mum$.
However, for these large grains the 9.7$\mum$ 
Si--O feature fades away 
(see Fig.\,6 of Greenberg 1996,
Fig.\,1 of Voshchinnikov \& Henning 2008).
Therefore, it is difficult to account for
the ``anomalous'' silicate emission features
of AGNs just in terms of an increase in grain size. 

A non-spherical grain shape or shape distribution
can also broaden the 9.7$\mum$ Si--O feature and
red-shift its peak wavelength (see Li 2008). 
But to account for the peak wavelengths of
$\lambda_{\rm peak}$\,$\simali$10--11.5$\mum$
observed in some AGNs, the dust has to be extremely 
elongated which is unrealistic
(e.g. spheroidal dust needs to have an elongation $>$6).

The peak wavelength of the 9.7$\mum$ Si--O feature 
also depends on the silicate mineralogy. But we are
not aware of any amorphous silicate species
that could give rise to a Si--O feature 
peaking at $\lambda >10\mum$.
 
  Crystalline olivine has its Si--O feature peaks at
  $\simali$11.2$\mum$ (see Yamamoto et al.\ 2008). 
  But this feature is too sharp 
  compared to the broad Si--O emission features of AGNs.
  Within the signal-to-noise ratio limits, we see
  no clear evidence for crystalline silicates in 3C\,273 
  and NGC\,3998 (see Fig.\,\ref{fig:sil_obs_spec}).
  So far, the detection of crystalline silicate dust
  in AGNs has only been reported in the BAL
  (broad absorption line) quasar PG\,2112+059  
  (Markwick-Kemper et al.\ 2007).\footnote{%
    Spoon et al.\ (2006) reported firstly the detection
    of narrow absorption features of crystalline silicates
    in ultraluminous IR galaxies (ULIRGs).
    
}
The 9.7$\mum$ Si--O feature emission features of
the protoplanetary disks around T Tauri and Herbig Ae/Be
stars are often also much broader than that of the ISM
(e.g. see Bouwman et al.\ 2001, Forrest et al.\ 2004).
This is usually interpreted as an indicator of grain 
growth (e.g. see Natta et al.\ 2007). 
The porous composite dust model was recently used
by Voshchinnikov \& Henning (2008) to demonstrate 
that a similar behavior of the feature shape occurs 
when the porosity of fluffy dust varies.
Kr\"ugel \& Siebenmorgen (1994) have also demonstrated
that a porous structure results in the broadening,
weakening and redshifting of the 9.7$\mum$ Si--O feature.
Fluffy dust aggregates are also naturally expected in 
protoplanetary disks as a result of grain coagulation.  

We plot in Figure \ref{fig:agn_lambdap_porosity}
as a function of porosity $P$
the peak wavelengths $\lambda_{\rm peak}$ of 
the 9.7$\mum$ Si--O feature of the absorption 
efficiency profiles $\Qabs(\lambda)$ 
and the emission profiles obtained by 
folding $\Qabs(\lambda)$ with the Planck function 
$B_\lambda(T)$ at various temperatures. 
It is seen that for 3C\,273 and NGC\,3998,
to account for the observed $\lambda_{\rm peak}\approx 10.6\mum$,
one requires either 
cool dust (with $T$\,$\simali$140--150$\K$)
with a low porosity ($P<0.3$),
or warm dust (with $T$\,$\simali$170--220$\K$)
with a high porosity ($P>0.6$).
The 18$\mum$ O--Si--O emission feature further
constrains that cool dust is preferred in 3C\,273
while NGC\,3998 appears to favour warm dust.

Also seen in Figure \ref{fig:agn_lambdap_porosity}
is that the range of the peak wavelengths 
$10\mum \simlt \lambda_{\rm peak}\simlt 11.5\mum$ 
of the 9.7$\mum$ Si--O emission features of 
the five PG quasars of Hao et al.\ (2005) falls 
within the predictions of
the porous composite dust model.
This indicates that the long-wavelength shifted 
9.7$\mum$ Si--O features of these quasars can be
accounted for by the combined effects of porosity
and temperature (i.e. either highly porous warm dust
or cold dust with a lower porosity).
The 18$\mum$ O--Si--O emission feature will allow us
to break this degeneracy.
Finally, for those AGNs whose 9.7$\mum$ Si--O absorption
or emission features do not exhibit any long-wavelength
shift, one may resort to dust with a low porosity
$P<0.3$ (and $T>220\K$ if they are in emission).
 
We should stress that although the combined effects
of porosity and temperature suffice by themselves 
to explain the general trend of broadening the 9.7$\mum$
Si--O emission features of AGNs and shifting their
peak wavelengths to longer wavelengths, 
we by no means exclude the effects of other factors 
such as grain size,\footnote{%
  The grain size effect is probably (at least in part)
  responsible for the nondetection of the 9.7$\mum$
  and 18$\mum$ silicate emission features in many
  type 1 AGNs (e.g. see Hao et al.\ 2007).
  The silicate size distribution in some AGNs might
  be dominated by large grains 
  ($>$ a few $\mum$; see Maiolino et al.\ 2001a,b)
  or small silicate grains are depleted
  (e.g. see Laor \& Draine 1993, Granato \& Danese 1994).
  Alternative explanations include sophisticated torus geometries
  (e.g. tapered disk configurations [Efstathiou \& Rowan-Robinson 1995],
   clumpy torus models [Nenkova et al.\ 2002, 2008;
   but see Dullemond \& van Bemmel 2005])
  and an assumption of strong anisotropy of 
  the source radiation (Manske et al.\ 1998).  
  }
shape, and mineralogy. It is very likely that all these
factors act together to produce the anomalous silicate 
emission features seen in AGNs.

Finally, we emphasize that although a steeply rising 
(cold) Planck function could redshift the 9.7$\mum$
Si--O emission feature, the observed shift of the peak
wavelength of this emission feature in AGNs cannot be 
purely a temperature effect since the silicate absorption 
profiles of some AGNs also appear anomalous;
e.g., the ``9.7$\mum$'' silicate feature of Mkn 231, 
a peculiar type 1 Seyfert galaxy, is seen in absorption
peaking at $\simali$10.5$\mum$ (Roche et al.\ 1983);
Jaffe et al.\ (2004) found that the 9.7$\mum$ silicate 
absorption spectrum of NGC\,1068 shows 
a relatively flat profile from 8 to 9$\mum$ 
and then a sharp drop between 9 and 10$\mum$;
in comparison, the Galactic silicate absorption profiles
begin to drop already at $\simali$8$\mum$.

To conclude, we have explored in a quantitative way
on the effects of grain porosity on the silicate Si--O
stretching feature using the multi-layered sphere model. 
It is found that the Si--O feature 
broadens and shifts to longer wavelengths 
with the increasing of dust porosity.
We conclude that the combined effects of dust porosity 
and cool temperature of $T<$\,200$\K$
(which further redshifts the silicate feature)
could explain the observed broadening and longer-wavelength 
shifting of the 9.7$\mum$ Si--O feature of AGNs.

\section*{Acknowledgments}
We thank L. Hao and E. Strum for providing us with
{\it Spitzer} IRS spectra of 3C\,273 and NGC\,3998.
We thank L. Hao and R. Siebenmorgen for 
very helpful comments.
ML and AL are supported in part by NASA/HST Theory
Programs, and NASA/Spitzer Theory Programs.
AL is supported by the NSFC
Outstanding Overseas Young Scholarship.

\bsp
\label{lastpage}


\begin{thebibliography}{}
%
\bibitem[]{}Antonucci, R.\ 1993, ARA\&A, 31, 473
\bibitem[]{}Bouwman, J., Meeus, G., de Koter, A., Hony, S., 
            Dominik, C., \& Waters, L.~B.~F.~M.\ 2001, A\&A, 375, 950
\bibitem[]{}Bowey, J.~E., Adamson, A.~J., \& Whittet, D.~C.~B.\ 
            1998, MNRAS, 298, 131 
\bibitem[]{}Dorschner, J., Begemann, B., Henning,
            J\"ager, C., Mutschke, H.\ 1995, A\&A, 300, 503
\bibitem[]{}Draine, B.~T.\ 1988, ApJ, 333, 848
\bibitem[]{}Draine, B.~T.\ 2003, ARA\&A, 41, 241
\bibitem[]{}Dullemond, C.~P., \& van Bemmel, I.~M.\ 
            2005, A\&A, 436, 47 
\bibitem[]{}Efstathiou, A.\ 2006, MNRAS, 371, L70 
\bibitem[]{}Efstathiou, A., \& Rowan-Robinson, M.\ 
            1995, MNRAS, 273, 649
\bibitem[]{}Forrest, W.~J., et al.\ 2004, ApJS, 154, 443
\bibitem[]{}Granato, G.~L., \& Danese, L.\ 1994, MNRAS, 268, 235
\bibitem[]{}Greenberg, J.~M.\ 1996
            in Cosmic Dust Connection, 
            ed. J.M. Greenberg
            (Dordrecht: Kluwer), 443
\bibitem[]{}Hao, L., et al.\ 2005, ApJ, 625, L75
\bibitem[]{}Hao, L., Weedman, D.~W., Spoon, H.~W.~W., 
            Marshall, J.~A., Levenson, N.~A., Elitzur, M.,
            \& Houck, J.~R.\ 2007, ApJ, 655, L77
\bibitem[]{}Jaffe, W., et al.\ 2004, Nature, 429, 47
\bibitem[]{}Kemper, F., Vriend, W.~J., Tielens, A.~G.~G.~M.\
            2004, ApJ, 609, 826 (Erratum: 2005, ApJ, 633, 534)
\bibitem[]{}Kimura, H., Mann, I., Biesecker, D.~A., 
            \& Jessberger, E.~K.\ 2002, Icarus, 159, 529 
\bibitem[]{}Kleinmann, D.E., Gillett, F.C., \& Wright, E.L.\ 
            1976, ApJ, 208, 42 
\bibitem[]{}Kr\"ugel, E., \& Siebenmorgen, R.\ 1994, 
            A\&A, 288, 929
\bibitem[]{}Laor, A., \& Draine, B.~T.\ 1993, ApJ, 402, 441
\bibitem[]{}Levenson, N.~A., et al.\ 2007, ApJ, 654, L45
\bibitem[]{}Li, A.\ 2007, in ASP Conf. Ser. 373,
            The Central Engine of Active Galactic Nuclei,
            ed. L. C. Ho \& J.-M. Wang (San Francisco: ASP), 561
\bibitem[]{}Li, A.\ 2008, in Small Bodies in Planetary Sciences
            (Lecture Notes in Physics Series), I. Mann, A. Nakamura,
            \& T. Mukai (eds.), Springer, 167
\bibitem[]{}Li, A., \& Lunine, J.~I.\ 2003, 590, 368
\bibitem[]{}Maiolino, R., Marconi, A., Salvati, M., Risaliti, G.,
            Severgnini, P., Oliva, E., La Franca, F.,
            \& Vanzi, L.\ 2001a, A\&A, 365, 28
\bibitem[]{}Maiolino, R., Marconi, A., \& Oliva, E.\
            2001b, A\&A, 365, 37
\bibitem[]{}Manske, V., Henning, Th., \& Men'shchikov, A.B.\ 
            1998, A\&A, 331, 52 
\bibitem[]{}Markwick-Kemper, F., Gallagher, S.~C., Hines, D.~C., 
            \& Bouwman, J.\ 2007, ApJ, 668, L107
\bibitem[]{}Marshall, J.A., Herter, T.L., Armus, L., Charmandaris, V., 
            Spoon, H.W.W., Bernard-Salas, J., \& Houck, J.R.\ 2007, 
            ApJ, 670, 129 
\bibitem[]{}Natta, A., Testi, L., Calvet, N., Henning, T., 
            Waters, R., \& Wilner, D.\ 2007, 
            in Protostars and Planets V, 
            ed. B. Reipurth, D. Jewitt, \& K. Keil 
            (Tucson: Univ. Arizona Press), 767
\bibitem[]{}Nenkova, M., Ivezic, Z., \& Elitzur, M.\ 
            2002, ApJ, 570, L9
\bibitem[]{}Nenkova, M., Sirocky, M.~M., Nikutta, R., Ivezic, Z., 
            \& Elitzur, M.\ 2008, ApJ, in press
\bibitem[]{}Rieke, G.H., \& Low, F.J.\ 1975, ApJ, 199, L13 
\bibitem[]{}Roche, P.F., Aitken, D.K., \& Whitmore, B.\ 1983, 
             MNRAS, 205, P21 
\bibitem[]{}Roche, P.~F., Aitken, D.~K., \& Smith, C.~H.\
            1991, MNRAS, 252, 282
\bibitem[]{}Roche, P.F., Packham, C., Aitken, D.K.,
             \& Mason, R.E.\ 2007, MNRAS, 375, 99
\bibitem[]{}Rouleau, F., \& Martin, P.~G.\ 1991, ApJ, 377, 526
\bibitem[]{}Schweitzer, M., et al.\ 2008, ApJ, 679, 101
\bibitem[]{}Shi, Y., et al.\ 2006, ApJ, 653, 127
\bibitem[]{}Siebenmorgen, R., Kr\"{u}gel, E.,
            \& Spoon, H.~W.~W.\ 2004, A\&A, 414, 123
\bibitem[]{}Siebenmorgen, R., Haas, M., Kr\"{u}gel, E.,
            \& Schulz, B.\ 2005, A\&A, 436, L5
\bibitem[]{}Spoon, H.W.W., et al.\ 2006, ApJ, 638, 759
\bibitem[]{}Spoon, H.~W.~W., Marshall, J.~ A., Houck, J.~R.,
            Elitzur, M., Hao, L., Armus, L., Brandl, B.~R.,
            \& Charmandaris, V.\ 2007, ApJ, 654, L49
\bibitem[]{}Sturm, E., et al.\ 2005, ApJ, 629, L21
\bibitem[]{}Sturm, E., Hasinger, G., Lehmann, I., 
              Mainieri, V., Genzel, R., Lehnert, M. D.,
              Lutz, D., \& Tacconi, L. J.\ 2006, ApJ, 642, 81
\bibitem[]{}Teplitz, H.I., et al.\ 2006, ApJ, 638, L1
\bibitem[]{}Urry, C.M., \& Padovani, P.\ 1995, PASP, 107, 803 
\bibitem[]{}Voshchinnikov, N.~V., \& Mathis, J.~S.\
            1999, ApJ, 526, 257
\bibitem[]{}Voshchinnikov, N.~V., Il'in, V.~B.,
            \& Henning, Th.\ 2005, A\&A, 429, 371
\bibitem[]{}Voshchinnikov, N.~V., Il'in, V.~B., Henning, Th.,
            \& Dubkova, D.~N.\ 2006, A\&A, 445, 167
\bibitem[]{}Voshchinnikov, N.~V., \& Henning, Th.\
            2008, A\&A, 483, L9
\bibitem[]{}Weedman, D.~W., et al.\ 2005, ApJ, 633, 706
\bibitem[]{}Wooden, D.~H., Harker, D.~E., Woodward, C.~E., Butner, H.~M.,
            Koike, C., Witteborn, F.~C., \& McMurtry, C.~W.\
            1999, ApJ, 517, 1034
\bibitem[]{}Yamamoto, T., Chigai, T., Kimura, H.,
            \& Tanaka, H.\ 2008, Earth, Planets \& Space, in press
\end{thebibliography}
\end{document}